\documentclass[copyright,creativecommons]{eptcs}
\providecommand{\event}{ML Family/OCaml Workshop 2015} % Name of the event you are submitting to
\usepackage{breakurl}             % Not needed if you use pdflatex only.

\title{Specialization of Generic Array Accesses After Inlining
  \break (System Description)}

\author{Ryohei Tokuda \qquad\qquad\qquad Eijiro Sumii \qquad\qquad\qquad Akinori Abe
\institute{Graduate School of Information Sciences, Tohoku University, Sendai, Japan}
 \email{tokuda@sf.ecei.tohoku.ac.jp \qquad sumii@ecei.tohoku.ac.jp \qquad abe@sf.ecei.tohoku.ac.jp}
}

\usepackage{amsmath}
\usepackage{listings}
\usepackage{graphicx}
\usepackage{url}
\usepackage{here}

\def\titlerunning{Specialization of Generic Array Accesses After Inlining}
\def\authorrunning{R. Tokuda, E. Sumii \& A. Abe}
\begin{document}
\maketitle

\begin{abstract}
  We have implemented an optimization that specializes
  type-generic array accesses after inlining of polymorphic functions in the native-code OCaml compiler.
  Polymorphic array operations (read and write) in OCaml require
  runtime type dispatch because of ad hoc memory representations of integer and float arrays.
  It cannot be removed even after being monomorphized by inlining
  because the intermediate language is mostly untyped.
  We therefore extended it with explicit type application like System F
  (while keeping implicit type abstraction by means of unique identifiers for type variables).
  % in the typed abstract syntax tree in OCaml
  Our optimization has achieved up to 21\% speed-up of numerical programs.
\end{abstract}

\section{Introduction}

\subsection{Background}

Representation of primitive values such as floating-point numbers
is a classical problem in the implementation of polymorphic languages
since ad hoc representations hinder uniform treatment of values.
The classical way to overcome this difficulty is
to fit every value in one machine word by
heap-allocating multi-word data (called \emph{boxing}) and manipulating them through pointers.
Although simple, this method is inefficient
especially for numerical computations because of
the heap allocation and pointer dereferences.
The cost of garbage collection
due to the frequent allocations is particularly problematic.

More sophisticated implementation methods for polymorphism have also been devised.
Leroy \cite{DBLP:conf/popl/Leroy92} and Shao \cite{DBLP:conf/icfp/Shao97}
adopt a mixture of specialized and uniform representations,
where the cost of conversions between different representations
(such as flat arrays vs. arrays of boxed values) is non-trivial.
Another method is to pass the types at runtime~\cite{DBLP:conf/popl/HarperM95,DBLP:journals/toplas/MorrisonDCB91,DBLP:conf/pldi/TarditiMCSHL96a},
where the construction of the type representations itself incurs an overhead.

OCaml takes an ad hoc approach to the problem:
it mainly adopts uniform representation
but uses unboxed representations for ``local'' floating-point
numbers (i.e., ones that do not escape a function's body)
and, in particular, for \emph{arrays} of floating-point numbers (as well as records with
fields of floating-point numbers only). Such specialized representation of
\lstinline|float array| means that polymorphic array accesses (such as
setting and getting the elements) need to
\emph{dynamically} check the type of the array and make a case branch
when it is a \lstinline|float array|.  These dynamic checks generally
incur runtime overheads, which the standard OCaml compiler tries to
remove if the monomorphic type of the array is statically known.%
\footnote{Standard
  ML (as well as OCaml's \lstinline|bigarray|) takes another ad hoc approach:
  its Basis Library offers a monomorphic
  module \lstinline|RealArray| (along with other monomorphic array
  modules of a common interface) for an unboxed representation.
  There is also a similar
  proposals~\cite{Frisch,lpw25} for \lstinline|array| in OCaml.}

% big-arrayではint32とint64の配列は同じようにフラットになっています

\subsection{Problem}

Despite the aforementioned specialization of generic array accesses,
the standard OCaml compiler fails to apply the specialization when the
monomorphic type is known \emph{after} inlining of functions,
because of a lack of type information in the intermediate language.
% \footnote{OCaml's intermediate language is
%   generally considered untyped, but this (partial) type information
%   for specialized generic array accesses is an exception.}

To see the problem concretely, consider the following program:
\begin{lstlisting}
let get0 a = a.(0)  (* 'a array -> 'a *)
let i = get0 int_array
let f = get0 float_array
\end{lstlisting}
Even if the polymorphic function \lstinline{get0} is inlined,
the array accesses \lstinline{int_array.(0)} and
\lstinline{float_array.(0)} are \emph{not} specialized and are
considered generic, incurring the runtime overheads of case branches
over the dynamic types of the (obviously monomorphic) arrays.

This problem is due to the internal representation of the partial type information
attached (only) to array accesses in the intermediate language, which is defined in the OCaml compiler as (roughly speaking):
\begin{lstlisting}
type array_kind =
  | Pgenarray   (* generic           *)
  | Pintarray   (* int               *)
  | Pfloatarray (* float             *)
  | Paddrarray  (* address (pointer) *)
\end{lstlisting}
% \lstinline|array_kind| annotates how to access an array to each array operations.
The \lstinline|array_kind| ``\lstinline|Pgenarray|'' means runtime dispatch over the dynamic type of the elements of an array.

For example, the program above can be annotated with this (partial)
type information like
\begin{lstlisting}
let get0 a = a.{Pgenarray}(0)
let i = get0 int_array
let f = get0 float_array
\end{lstlisting}
where \lstinline|{}| denotes the internal type annotation with \lstinline|array_kind|.
Obviously, just inlining the function \lstinline|get0| does not specialize
the generic read operations:
% because all polymorphic type are translated to just \lstinline|Pgenarray|
\begin{lstlisting}
let i = int_array.{Pgenarray}(0)
let f = float_array.{Pgenarray}(0)
\end{lstlisting}

\subsection{Key Ideas}

% スライドの通り進行する

Our idea is to add explicit type information like System F
to the mostly untyped intermediate language \lstinline|lambda| of OCaml.
That is, we basically extend the intermediate language with
type abstractions and applications.

For instance, the example above can be type-annotated like:
\begin{lstlisting}
let get0{'a} a = a.{'a}(0)
let i = get0{Pintarray} int_array
let f = get0{Pfloatarray} float_array
\end{lstlisting}
We add the formal type parameter \lstinline|{'a}| in the definition of
\lstinline|get0|. % to take an \lstinline|array_kind|
Then, we attach the type information \lstinline|{'a}| on the read access \lstinline|a.(0)| to
the array \lstinline|a| of polymorphic type \lstinline|'a array|.
% We replace \lstinline|Pgenarray| with \lstinline|'a| because \lstinline|Pgenarray|
% is too abstract to be substituted with a concrete \lstinline|array_kind|.
We then explicitly denote type applications by
annotating \lstinline|get0| with \lstinline|{Pintarray}| and \lstinline|{Pfloatarray}|.
The generic array accesses can now be specialized by inlining as:
\begin{lstlisting}
let i = int_array.{Pintarray}(0)
let f = float_array.{Pfloatarray}(0)
\end{lstlisting}

For type application, we indeed extended the intermediate language.
For abstraction, we actually used globally unique identifiers for type variables
to avoid introducing a new binder, as OCaml does for typing; in exchange, we have to specify
\emph{which} type variable to instantiate at the application side.

% The IL is basically untyped except \lstinline|array_kind|, so
% such information is not necessary except the dynamic dispatches.
% We solve the problem by the following:

% \begin{itemize}
% \item type abstraction: use unique global ids (as usual in OCaml).
% \item type application: extend the IL (shown in next section).
% \end{itemize}

% Type variables are identified by its unique global id in OCaml compiler.
% This means the id (representing \lstinline|'a|) is only used to
% represent the same type variable. Then, no conflict happens if we
% replace \lstinline|'a| with a concrete \lstinline|array_kind| with no attention.

For example, the foregoing program is now represented like:
\begin{lstlisting}
let get0 (* {'a} *) a = a.{'a}(0)
let i = get0{'a$\mapsto$I} int_array
let f = get0{'a$\mapsto$F} float_array
\end{lstlisting}
%
% 上のプログラムの説明を丁寧にする
%
In the definition of \lstinline|get0|,
we omit the type abstraction \lstinline|(* {'a} *)|
while annotating the read access with the implicitly bound type variable \lstinline|'a|.
We then annotate the applications of \lstinline|get0| with
a mapping from the type variable \lstinline|'a| to an \lstinline|array_kind|,
like \lstinline|{'a$\mapsto$I}| and \lstinline|{'a$\mapsto$F}|.
When inlining \lstinline|get0|, we replace the occurrence of \lstinline|'a|
according to the given mapping, resulting in
\begin{lstlisting}
let i = int_array.{I}(0)
let f = float_array.{F}(0)
\end{lstlisting}
as desired.

Of course, not all polymorphic functions can be inlined completely
(because of code size, for example).  In such cases, some type
variables still remain and are compiled as generic array accesses with dynamic
type dispatch.

\section{Implementation}

In this section, we describe details of our implementation based on the native-code OCaml compiler.  We have made the following changes to the intermediate language \lstinline|lambda| (resp.~\lstinline|clambda|) before (resp.~after) closure conversion.

\subsection{Refining the type information}
First, we replace \lstinline|Pgenarray| (type of generic arrays) with \lstinline|Ptvar of int|
(type variable with an integer identifier) in \lstinline|array_kind|
to specify \emph{which} type variable the generic type refers to (like \lstinline|{'a}| in \lstinline|a.{'a}(0)| instead of \lstinline|Pgenarray| in \lstinline|a.{Pgenarray}(0)|) as follows%
\footnote{Precisely speaking, we still keep \lstinline|Pgenarray|
for places where our current implementation abandons specialization, such as functors and GADTs.}:

\begin{lstlisting}
type array_kind =
  Ptvar of int (* id *) | Pintarray | Pfloatarray | Paddrarray
\end{lstlisting}

\subsection{Making type applications explicit}
% 次にLspecializedをlambda言語に追加する
% (Lspecializeのパラメータを説明する)

Second, we add \lstinline|Lspecialized| to the intermediate languages for explicitly representing type applications.

\begin{lstlisting}
type lambda = (* ditto for clambda *)
| ...(* same as before *)...
| Lspecialized of lambda * kind_map (* type application *)
and kind_map = (int * array_kind) list (* association list *)
\end{lstlisting}

The constructor \lstinline|Lspecialized| has two parameters: the first parameter (of type \lstinline|lambda|) is
a polymorphic function to be specialized,
while the second (type \lstinline|kind_map|) is a type mapping described above,
such as \lstinline|{'a$\mapsto$I}| and \lstinline|{'a$\mapsto$F}|.
We insert \lstinline|Lspecialized| to every occurrence of a let-bound polymorphic variable
during the translation from \lstinline|typedtree| (typed AST) to \lstinline|lambda|.%
\footnote{Thus, the only possible first parameter for \lstinline|Lspecialized| is actually
an occurrence of a local (\lstinline|Lvar|) or global (module access) variable, so it is also possible to attach the \lstinline|kind_map| to those variable occurrences of instead of introducing \lstinline|Lspecialized|.  We did not take this approach because modifying the module access primitive (\lstinline|Pgetglobal|) seemed more complicated than simply adding \lstinline|Lspecialized|.}
% Lspecializedの要素であるkind_mapの作り方は、
% 型環境に入っている多相的な型とTyped ASTに入っている単相型を見比べて
% 型がどのように単相化されたのか復元する
% OriginalのUnify関数は副作用で書き換えるものしかないので
% Pureな見比べるUnify関数を自作した
More specifically, for every variable occurrence,
we compare the monomorphic type of the variable (annotated in the AST)
with the polymorphic type stored in the type environment
and, if the latter type is indeed polymorphic, insert \lstinline|Lspecialized|
with \lstinline|kind_map| recovered from the comparison
by means of a one-directional unification.
(In principle, this mapping is already known during type inference
but is discarded in the current OCaml compiler.  We avoided modifying
the type inference because of its complexity.)

% 型変数のα変換をすべて追う(ファイルをまたいだケース)

\subsection{Recording and following the renaming of type variables}
Finally, we need to record and follow renaming of type variables exported via \lstinline|.cmx| files
to prevent inconsistencies with \lstinline|.cmi| files.

Suppose, for example, that we compile a source file \lstinline|a.ml|
including function \lstinline|get0|
\begin{lstlisting}[title=\lstinline|a.ml|]
let get0 a = a.{'a}(0)
\end{lstlisting}
with the interface file:
\begin{lstlisting}[title=\lstinline|a.mli|]
val get0 : 'a array -> 'a
\end{lstlisting}
On one hand, the \lstinline|.mli| is compiled into a \lstinline|.cmi| file:
\begin{lstlisting}[title=\lstinline|a.cmi|]
(* pseudo-code for the binary *)
val get0 : 'a array -> 'a
\end{lstlisting}
On the other hand, the implementation \lstinline|a.ml| is compiled
separately from the interface \lstinline|a.mli|.  Even the types are
inferred independently---they are only \emph{checked} against the
compiled interface \lstinline|a.cmi| \emph{after} inference.  As a
result, the type variable \lstinline|'a| may be given a completely
different identifier in the AST \lstinline|a.cmx| generated for inlining:
\begin{lstlisting}[title=\lstinline|a.cmx|]
(* pseudo-code for the AST *)
get0 a = a.{'b}(0)
\end{lstlisting}
This inconsistency is problematic for our scheme, where
\lstinline|get0| is applied like \lstinline|A.get0{'a$\mapsto$I}int_array|
according to its type in the interface \lstinline|a.cmi|.  We fixed this by adjusting the type
variable identifiers in a \lstinline|.cmx| according to the
corresponding \lstinline|.cmi| just before generating the former:
\begin{lstlisting}[title=\lstinline|a.cmx| (adjusted)]
(* pseudo-code for the AST *)
get0 a = a.{'a}(0)
\end{lstlisting}

Moreover, suppose that the function \lstinline|A.get0| is used from
another file \lstinline|b.ml|.  When the interface
\lstinline|a.cmi| is imported, the type variable \lstinline|'a| is renamed
for the sake of uniqueness inside the importing module \lstinline|B|
\begin{lstlisting}[title=\lstinline|b.ml|]
open A
(* val get0 : 'c array -> 'c *)

let i = get0{'c$\mapsto$I} int_array (* an example using A.get0 *)
\end{lstlisting}
and becomes inconsistent with the implementation \lstinline|a.cmx|.
We fixed this inconsistency by remembering the renaming such as
\lstinline|'a$\mapsto$'c| in a global table at import time, and applying it before
inlining function bodies such as \lstinline|a.{'a}(0)|.
\begin{lstlisting}[title=\lstinline|b.ml| (after inlining)]
open A
(* renaming table for A is 'a$\mapsto$'c *)
(* val get0 : 'c array -> 'c *)

let i = int_array.{('c$\mapsto$I)('a$\mapsto$'c)'a}(0) (* correctly inlined *)
\end{lstlisting}

% There is a more complex case: define an alias of \lstinline|A.get0|.

% \begin{lstlisting}[caption=c.ml]
% (* $\alpha$-renaming table: {'b$\mapsto$'c} *)
% (* val A.get0: 'c array->'c *)

% (* val f : 'd array -> 'd *)
% let f = A.get0{'c -> 'd}
% (* val g : 'e array -> 'e *)
% let g = f{'d -> 'e}
% (* val h : 'f array -> 'f *)
% let h = f{'d -> 'f}

% let x = g{'e -> I} int_array
% \end{lstlisting}

% To specialize \lstinline|g|, we follow alpha renaming repeatedly.

\section{Experiments}
\label{sec:exp}
% 何を追加するべき？

We have implemented the above specialization\footnote{Our compiler is available from: \url{https://github.com/nomaddo/ocaml}} on top of
the 4.02 branch of OCaml as of May 6, 2015,
and measured its effects
for the numerical programs in Table~1.  ``Simple'' is a program that
adds all elements of an array, where all the accesses are made through
polymorphic functions.  ``Random'' makes generic array accesses,
randomly switching between \lstinline|int array| and %
\lstinline|float array| (in addition, the pseudo-random number
generator internally makes monomorphic array accesses).  The other
benchmarks are realistic, naturally written numerical programs: ``DKA''
stands for Durand-Kerner-Aberth (a method for finding a root of a
complex polynomial), ``FFT'' is a fast Fourier transform, ``K-means'' is a
clustering method used for data mining and machine learning, ``LD'' stands
for the Levinson-Durbin recursion for time series analysis, ``LU'' is the
LU decomposition in linear algebra, ``NN'' is a neural network program,
and ``QR'' is the QR decomposition (again in linear algebra).  We have
hand-tuned the timing of garbage collections.  We compiled all the
files (including the standard library modules) with %
\lstinline|ocamlopt -inline 10000000| (and \lstinline|-unsafe| for the
benchmark programs\footnote{Their source code is available from:
\url{https://github.com/nomaddo/ocaml-numerical-analysis/tree/bench}}).

% 表をスライドのものに合わせた

\begin{table}[htb]
  \centering
  \begin{tabular}{l}
  \begin{tabular}{lrrrrr} \hline
    Program & \multicolumn{1}{c}{All}       & \multicolumn{2}{c}{Gen-before} & \multicolumn{2}{c}{Gen-after} \\
    \hline
    Simple  & 219,999,999  & 219,999,999  & (100.0\%) & 0          & (0.0\%)  \\
    Random  & 1,200,000,001 & 599,999,999  & (50.0\%)  & 0          & (0.0\%)  \\
    DKA     & 38,436,359   & 12,899,744   & (33.6\%)  & 1,748       & (0.0\%)  \\
    FFT     & 222,298,106  & 6,291,450    & (2.8\%)   & 0          & (0.0\%)  \\
    K-means & 16,996,284   & 13,022,928   & (76.6\%)  & 11,562,804   & (68.0\%) \\
    LD      & 1,583,266,345 & 250,025,000  & (15.8\%)  & 0          & (0.0\%)  \\
    LU      & 555,548,601  & 7,603,452    & (1.4\%)   & 7,589,706    & (1.4\%)  \\
    NN      & 1,324,325,917 & 671,112,411  & (50.7\%)  & 368,199,383  & (27.8\%) \\
    QR      & 577,566,417  & 571,855,214  & (99.0\%)  & 133,758     & (0.0\%)  \\
    \hline
  \end{tabular}
  \\ \\
  \begin{tabular}{lrrr} \hline
    Program & Time-before & Time-after & Speed-up \\
    \hline
    Simple   & 0.372 s  & 0.354 s & 5\% \\
    Random   & 4.619 s  & 3.939 s & 15\% \\
    DKA      & 0.370 s  & 0.363 s & 2\% \\
    FFT      & 6.119 s  & 6.136 s & 0\% \\
    K-means  & 0.406 s  & 0.401 s & 1\% \\
    LD       & 7.324 s  & 7.115 s & 3\% \\
    LU       & 3.261 s  & 3.263 s & 0\% \\
    NN       & 12.20 s  & 11.75 s & 4\% \\
    QR       & 6.255 s  & 4.931 s & 21\% \\
    \hline
  \end{tabular}
  \end{tabular}
  \caption{Results of experiments: ``All'' is the number of all array accesses, and ``Gen-before'' (resp.~``Gen-after'') is generic array accesses before (resp.~after) the specialization, and ``Speed-up'' is $(\textrm{Time-after} - \textrm{Time-before}) / \textrm{Time-before} \times 100\%$.}
\end{table}

% \begin{table}[tp]
%   \centering
%   \begin{tabular}{|l|r|rr|rr|r|} \hline
%     Program & All & \multicolumn{2}{r|}{Gen (before)} & \multicolumn{2}{r|}{Gen (after)}  & Speed-up \\
%     \hline
%     Simple  & 219999999  & 219999999 & (100.0\%) & 0 & (0.0\%)          & 5\% \\
%     Random  & 1200000001 & 599999999 & (50.0\%)  & 0 & (0.0\%)          & 15\% \\
%     DKA     & 38436359   & 12899744 & (33.6\%)   & 1748 & (0.0\%)       & 2\% \\
%     FFT     & 222298106  & 6291450 & (2.8\%)     & 0 & (0.0\%)          & 0\% \\
%     K-means & 16996284   & 13022928 & (76.6\%)   & 11562804 & (68.0\%)  & 1\% \\
%     LD      & 1583266345 & 250025000 & (15.8\%)  & 0 & (0.0\%)          & 3\% \\
%     LU      & 555548601  & 7603452 & (1.4\%)     & 7589706 & (1.4\%)    & 0\% \\
%     NN      & 1324325917 & 671112411 & (50.7\%)  & 368199383 & (27.8\%) & 4\% \\
%     QR      & 577566417  & 571855214 & (99.0\%)  & 133758 & (0.0\%)     & 21\% \\
%     \hline
%   \end{tabular}
%   \caption{Results of experiments: ``All'' is the number of all array accesses, ``Gen (before)'' (resp.~``Gen (after)'') is generic array accesses before (resp.~after) the specialization, and ``Speed-up'' is $(\textrm{Time (after)} - \textrm{Time (before)}) / \textrm{Time (before)} \times 100\%$.}
% \end{table}

The results (on Ubuntu Linux 14.04, Intel(R) Core(TM) i7-2677M 1.80
GHz---fixed at 1.0 GHz by \lstinline|cpufreq| to avoid experimental errors
caused by Intel Turbo Boost, which is sensitive to changes in temperature!---and 4 GiB DDR3 SDRAM) are also in
Table~1 and can be explained as follows:
\begin{itemize}
\item ``Simple'', ``DKA'', and ``LD'' show modest speed-up because all the
  generic accesses are specialized while most of the execution time is
  still spent on floating-point operations.
\item ``Random'' exhibits considerable improvement since the generic
  accesses are removed and the program does not perform any other
  significant computation.
\item ``FFT'' and ``LU'' contain only a relatively small number of generic accesses in the
  first place because of the monomorphic coding style.
\item ``K-means'', ``NN'', and ``LU'' include polymorphic array access
  functions that are not inlined at all, probably because of
  closure sharing of the OCaml compiler.  For instance, the
  following function \lstinline|foldi|
\begin{lstlisting}
let foldi f init x =
  snd (Array.fold_left
         (fun (i, acc) xi -> (i+1, f i acc xi))
         (0, init)
         x)
\end{lstlisting}
  is never inlined, since an argument \lstinline|f| of
  \lstinline|foldi| appears free in the anonymous function \\
  \lstinline|fun (i, acc) xi -> (i+1, f i acc xi)|.
\item The high-level coding style of QR almost exclusively uses
  polymorphic functions such as \lstinline|Array.map| and
  \lstinline|Array.fold_left| as opposed to low-level index accesses,
  and achieves 21\% speed-up thanks to the almost complete
  specialization.
\end{itemize}

\section{Related Work}

% Type-passingな研究のインライン展開がSpedizliationになる
% 型をちゃんと全部作ってわたすコストが大きいが、
% OCamlの型情報は部分的なので大した問題にならない
% しかし我々の仕事はコンパイル時間の話なので関係ないと思う

% 90年代くらいに散々研究された型付き中間言語の研究を並べる

% The original problem comes from the lack of type information in the intermediate language;
% type information is discarded during the translation from
% the typed AST to the intermediate language \lstinline|lambda|.
% If OCaml preserved types and adopted System F like intermediate language,
% no tricky hacks would not be required.

While we have extended the partial explicit type information for array
access operations in the intermediate language(s) of OCaml,
more explicitly typed intermediate languages have already been studied extensively:
\begin{itemize}
\item Harper and Morrisett~\cite{DBLP:conf/popl/HarperM95} formalized a translation
from the implicitly typed ML core language to
an explicitly typed intermediate language $\lambda_i^{ML}$,
which is a variant of $F_\omega$ extended with intensional type analysis.
% In this language, types annotates pair operations
% (its creation and projection) and applications of functions
% to some optimization (flattening tuples and arguments, use of specialized representations, etc).
Crary and Weirich~\cite{DBLP:conf/icfp/CraryW99} further extended $\lambda_i^{ML}$,
generalizing the type language.

\item TIL~\cite{DBLP:conf/pldi/TarditiMCSHL96a},
a Standard ML compiler with intensional polymorphism,
adopted $\lambda_i^{ML}$ as the intermediate language for type-directed optimizations
and applied conventional optimizations
(inlining, uncurrying, common sub-expression elimination, etc)
to the type-level language to reduce its overheads, in particular the construction of type representations.

\item FLINT~\cite{DBLP:journals/sigsoft/Shao00,DBLP:conf/icfp/ShaoLM98}, the
intermediate language of SML/NJ~\cite{smlnj}, used directed acyclic
graphs instead of trees as type representations for scalability
against large types.

% ML-KIT adopted System $F_\omega$ like intermediate language
% TODO: 論文にはそう書いてあるけどコード見ても確認できない。
% ML-KITの中間言語の論文も見つからない

\item GHC uses System $F_C$\cite{DBLP:conf/tldi/SulzmannCJD07},
an extension of $F_\omega$ with type coercion for uniformly supporting a wide variety of features
such as GADTs~\cite{DBLP:conf/popl/XiCC03} and
associated types~\cite{DBLP:conf/popl/ChakravartyKJM05, DBLP:conf/icfp/ChakravartyKJ05}.
\end{itemize}

Compared with these fully typed intermediate languages, our extension
(\lstinline|Ptvar| and \lstinline|Lspecialized|) to the partial type
information (\lstinline|array_kind|) for array access operations in OCaml is ad
hoc but small and relatively easy, with less than 1000 lines of modification
to (hundreds of thousands lines of) the native-code compiler.

\section{Conclusion}

We have made a relatively simple modification to the native-code OCaml
compiler to specialize generic array accesses after
inlining, and observed modest or significant speed-ups for numerical
programs.

Currently, a new intermediate language \lstinline|flambda|~\cite{flambda} is under
development (independently of our work) in the \lstinline|trunk| branch of OCaml.  We expect that
it solves problems of the current \lstinline|lambda| (like closure
sharing hinders inlining as observed in Section~\ref{sec:exp}) as well as
making our approach even more effective by enabling the specialization
of recursive functions (which are never inlined by the current OCaml compiler), for example.

Although our experiments focused on the efficiency of floating-point
programs, the optimization may also be effective for generic functions
(such as \lstinline|Array.map| and \lstinline|Array.fold_left|)
applied to integer or pointer arrays.
It would also be interesting future work to adapt an approach similar
to ours for specializing other operations than array accesses, such as
polymorphic comparisons and unboxing local variables.

\section*{Acknowledgments}

We thank Jacques Garrigue for his help on hacking the OCaml compiler and the anonymous reviewers for valuable comments.
This work was partially supported by JSPS KAKENHI Grant Numbers JP22300005, JP25540001, JP15H02681, JP16K12409, and by Mitsubishi
Foundation Research Grants in the Natural Sciences.
% We recommend abbrvnat bibliography style.

\nocite{*}
\bibliographystyle{eptcs}
\bibliography{refs}
\end{document}